\documentclass[prd,preprint,superscriptaddress,showpacs,byrevtex]{revtex4}

\usepackage{epsfig}
\newcommand{\be}{\begin{equation}}
\newcommand{\ee}{\end{equation}}
\newcommand{\bea}{\begin{eqnarray}}
\newcommand{\eea}{\end{eqnarray}}

\begin{document}

\title{$\chi_{cJ}$ Polarization in Polarized Proton-Proton Collisions at RHIC}

\author{Gouranga C. Nayak}
\affiliation{22 West Fourth Street \#1, Lewistown, Pennsylvania 17044, USA }
\date{\today}
\begin{abstract}

We study inclusive $\chi_{cJ}$ production with definite polarizations in polarized
proton-proton collisions at $\sqrt{s}$ = 200 GeV and 500 GeV at RHIC by using
non-relativistic QCD (NRQCD) color-octet mechanism. We present results of
rapidity distribution of $\chi_{c0}$, $\chi_{c1}$ and $\chi_{c2}$ production
with specific polarizations in polarized p-p collisions at RHIC within the PHENIX
detector acceptance range. We also present the corresponding results for the spin
asymmetries.

\end{abstract}

\maketitle
%\narrowtext
\newpage
\section{Introduction}

RHIC (relativistic heavy ion collider) at BNL studies
quark-gluon plasma in heavy-ion collisions and spin
structure of the proton in polarized p-p collisions. The spin program
at RHIC involves polarized proton collisions at $\sqrt{s}$=200 GeV and
500 GeV \cite{spin}. The quark-gluon plasma program involves Au-Au collisions
at $\sqrt{s}$=200 GeV \cite{qgp}.

Measurements of heavy probes such as $J/\psi$, $\psi'$ and $\chi_{cJ}$
are useful tools to detect quark-gluon plasma in heavy ion collisions
and to extract polarized gluon distribution function inside proton in
polarized p-p collisions \cite{phenix,QWGBrambilla}. Hence it is necessary
to analyze heavy quarkonium production mechanism in polarized p-p collisions at
RHIC. Non-relativistic QCD (NRQCD) color-octet mechanism \cite{bodwin,nayak}
has been successful to study heavy quarkonium production at high
energy colliders and at fixed target experiments.

The energy eigenstates of heavy quarkonium bound states $|H>$ in NRQCD
are labelled by the quantum numbers $J^{PC}$, with an additional
superscript to give the color; (1) for singlet and (8) for octet.
In the expansion of the Fock states the dominant component
in S-wave orthoquarkonium is the pure quark-antiquark state
$|Q\bar Q[^3S^{(1)}_1]>$. The state, such as
$|Q\bar Q[^3P^{(8)}_J]g>$ with dynamical gluons contribute with a probability
of order $v^2$,
where $v$ is the typical velocity of the non-relativistic heavy quark
(and antiquark), while the other states, such as
$|Q\bar Q[^3S^{(1,8)}_1]gg>$, $|Q\bar Q[^1S^{(8)}_0]g>$ and
$|Q\bar Q[^3D^{(1,8)}_J]gg>$
contribute to the probability in higher orders in $v$.
In P-wave orthoquarkonia, the dominant states are
$|Q\bar Q[^3P^{(1)}_J]>$ with the states having dynamical gluons such as $|Q\bar Q[^3S^{(8)}_1]g>$
contribute with a probability of order $v^2$.
Once the $Q\bar Q$ is formed in a color octet state it may emit a soft gluon
to transform into the color singlet state
$|Q\bar Q[^3P^{(1)}_J]>$ and become a $J/\psi$ by photon decay.
Also the $Q\bar Q$ pair in a color octet state can emit two long
wavelength gluons to become $J/\psi$.
These low energy interactions are negligible and the
non-perturbative matrix elements, labelled by the above quantum numbers,
can be extracted from experiments or can be calculated using lattice field theory.

NRQCD mechanism for heavy quarkonia production
has been very successful in explaining data at high energy colliders
such as in the p-p collisions at LHC \cite{lhc}, in the p-$\bar {\rm p}$
collisions at Tevatron
\cite{tevatron,CDFoctet,CL,kniehl}, in the e-p collisions at HERA \cite{HERAoctet}, in the
e${}^+$-e${}^-$ collisions at LEP \cite{LEPoctet}
and also at fixed target experiments \cite{fixedtargetoctet}.
The PHENIX data for $J/\psi$ production
in unpolarized p-p collisions can be
explained by the NRQCD color octet mechanism \cite{cooper}.

In addition to unpolarized p-p collisions, RHIC offers a wide variety of measurements
with respect to $J/\psi$ production.
They involve $J/\psi$ production (with and without definite
polarizations) in unpolarized p-p, d-Au, Cu-Cu and Au-Au collisions
and in polarized p-p collisions. The
parton fragmentation contribution to heavy quarkonium production
will be very small at RHIC because the maximum transverse momentum of heavy quarkonium that can be measured
at the RHIC is around 10 GeV/c. Hence we will neglect the parton fragmentation contribution to heavy quarkonium
production in our study. We will focus on the main contributions to heavy quarkonium
production at the RHIC which are from the parton fusion processes \cite{cooper,pol}.

In unpolarized and polarized partonic collisions,
the inclusive heavy quarkonium production cross section
(summed over quarkonium polarization states) were calculated
in \cite{CL,fm} and \cite{pol,gm} respectively.
Similarly, the heavy quarkonium production cross sections with definite polarizations in
unpolarized partonic collisions were calculated
in \cite{braaten,CDFpolarization}. At RHIC, the inclusive
$j/\psi$ and $\psi'$ production with specific polarizations in polarized p-p
collisions was studied in the PHENIX detector acceptance range in \cite{jacknayak}.
Similarly the double spin asymmetries in P-wave charmonium hadroproduction
was considered for the first time in \cite{nowak}. For earlier studies in the similar 
direction see \cite{earlier}.

In this paper we will study $\chi_{c0}$, $\chi_{c1}$ and $\chi_{c2}$ polarizations
in polarized p-p collisions at $\sqrt{s}$=200 GeV and 500 GeV at RHIC.
The PHENIX collaborations at RHIC will study $\chi_{c0}$, $\chi_{c1}$ and $\chi_{c2}$ polarizations
in polarized p-p collisions at $\sqrt{s}$=200 GeV and 500 GeV \cite{phenix}.

We will evaluate the partonic level cross sections
for the processes $q \bar q, ~gg ~\rightarrow ~
\chi_{cJ}(\lambda)$ in polarized p-p collisions where
$\lambda$ is the helicity (polarization) of the heavy quarkonium state.
Our LO analysis considers only the $\chi_{cJ}(\lambda)$ production in the forward direction at a finite rapidity.
The reason we need these results is that the PHENIX collaboration at the
RHIC will measure $\chi_{cJ}$ production with definite
polarizations in polarized p-p collisions at $\sqrt s$ = 200 GeV and 500 GeV.
Since polarized heavy quarkonium production at the
Tevatron energy scale \cite{expt} is not explained by the NRQCD color-octet
mechanism \cite{CDFpolarization} it will be useful to compare our results for
$\chi_{cJ}$ polarizations with the future data at the RHIC.
The study of polarized heavy quarkonium
production in polarized p-p collisions at the RHIC is also unique in the
sense that it probes the spin transfer processes in perturbative QCD (pQCD).

Note that the spin projection method \cite{CL} is normally used to evaluate the inclusive cross section for
heavy quarkonium production (summed over polarization states)
in parton fusion processes. However, the
heavy quarkonium production cross section with specific polarization in the final state
can involve additional matrix elements that do not contribute when the
polarization is summed. This involves interference terms between
partonic processes that produce heavy quark-antiquark
pairs with different total angular momenta. These interference terms cancel upon summing over polarizations.
Such interference terms can be calculated by using the helicity decomposition
method \cite{braaten}. For this reason we will use the helicity decomposition
method to calculate the square of the matrix elements
for heavy quarkonium production
with definite helicity in polarized partonic collisions. After evaluating the partonic
level cross sections we will compute the rapidity distributions of the cross sections
and spin asymmetries of $\chi_{c0}$, $\chi_{c1}$ and $\chi_{c2}$ production with definite helicity
states in polarized p-p collisions at RHIC at $\sqrt s$ = 200 GeV and 500 GeV
within the PHENIX detector acceptance ranges.

The paper is organized as follows. In section II we derive the partonic
level cross sections for $\chi_{cJ}$ production with definite
polarizations in polarized q-$\bar{\rm q}$ and g-g parton fusion processes
using the helicity decomposition method within the NRQCD color-octet mechanism.
In section III we present the results for the differential rapidity distributions
and spin asymmetries for the $\chi_{cJ}(\lambda)$
in the PHENIX detector acceptance range in polarized p-p collisions
at $\sqrt s$ = 200 GeV and 500 GeV. We conclude in section IV.

\section{Inclusive $\chi_c$ Production with Definite
Helicities in Polarized Partonic Collisions}

In this section we will use the NRQCD color-octet mechanism
and derive the square of the matrix element
for inclusive $\chi_{cJ}$ production with definite helicities
in polarized partonic fusion processes.
We will consider the (polarized) partonic fusion processes
$q \bar q~\rightarrow ~ \chi_{cJ}(\lambda)$ and $gg ~\rightarrow ~ \chi_{cJ}(\lambda)$
where $\lambda$ is the helicity of the produced heavy quarkonium state $H(\lambda)$.
The helicity $\lambda= 0, \pm 1$ correspond to longitudinal and transverse
polarization states respectively.
As mentioned above, we will use the helicity decomposition method \cite{braaten}
within the NRQCD color-octet mechanism to calculate these processes where
both initial and final state particles are polarized.

\subsection{The $q\bar q$ fusion process}

At the amplitude level the matrix element for the light quark-antiquark ($q\bar q$) fusion
process $q(k_1) + \bar q(k_2) \rightarrow Q(p_1) + \bar Q(p_2)$
producing a heavy quark-antiquark ($Q\bar Q$) pair is given by
\bea
M_{q \bar q \rightarrow Q\bar Q}~=~\frac{g^2}{P^2}~ \bar{v}(k_2) \gamma_\mu T^a
u(k_1) \, \bar{u}(p_1) \gamma^\mu T^a v(p_2)\,,
\label{eq1}
\eea
where $P^\mu~=~p_1^\mu~+~p_2^\mu~=~k_1^\mu~+~k_2^\mu$ is the CM momentum of the pair
and $p_1^\mu=P^\mu/2+L^\mu_jq^j$ and $p_2^\mu=P^\mu/2-L^\mu_jq^j$ with $q^i$ being their
relative momentum in the CM frame and $L^\mu_j$ is the boost matrix defined in \cite{braaten}
with both Lorentz and three vector indices. Using the
non-relativistic heavy quark Pauli spinors ($\xi$ and $\eta$)
we obtain (up to terms linear in $q$):
\bea
|M_{q \bar q \rightarrow Q\bar Q}|^2~=~\frac{g^4}{4m^2}~
{\eta^\prime}^{\dagger} \sigma^i T^a {\xi^\prime} \,L^\mu_i\,
\bar{u}(k_1) \gamma_\mu T^a
v(k_2)
\,\bar{v}(k_2) \gamma_\nu T^b
u(k_1)\,
L^\nu_j \,{\xi}^{\dagger} \sigma^j T^b {\eta}\,,
\label{eq4}
\eea
where $m$ is the mass of the heavy quark. We consider incoming (massless) light quarks and
antiquarks
\bea
&& u(k_1) \bar{u}(k_1) ~=~ \frac{1}{2}~(1+ h_1\gamma_5)~
\gamma_\mu k_1^\mu
\nonumber \\
&&~v(k_2) \bar{v}(k_2) ~=~ \frac{1}{2}~(1- h_2\gamma_5)~
\gamma_\mu k_2^\mu \,
\label{hel}
\eea

\begin{figure}[htb]
\vspace{2pt}
%\centering{\rotatebox{270}{\epsfig{figure=higgspm2.ps,height=7cm}}}
\centering{{\epsfig{figure=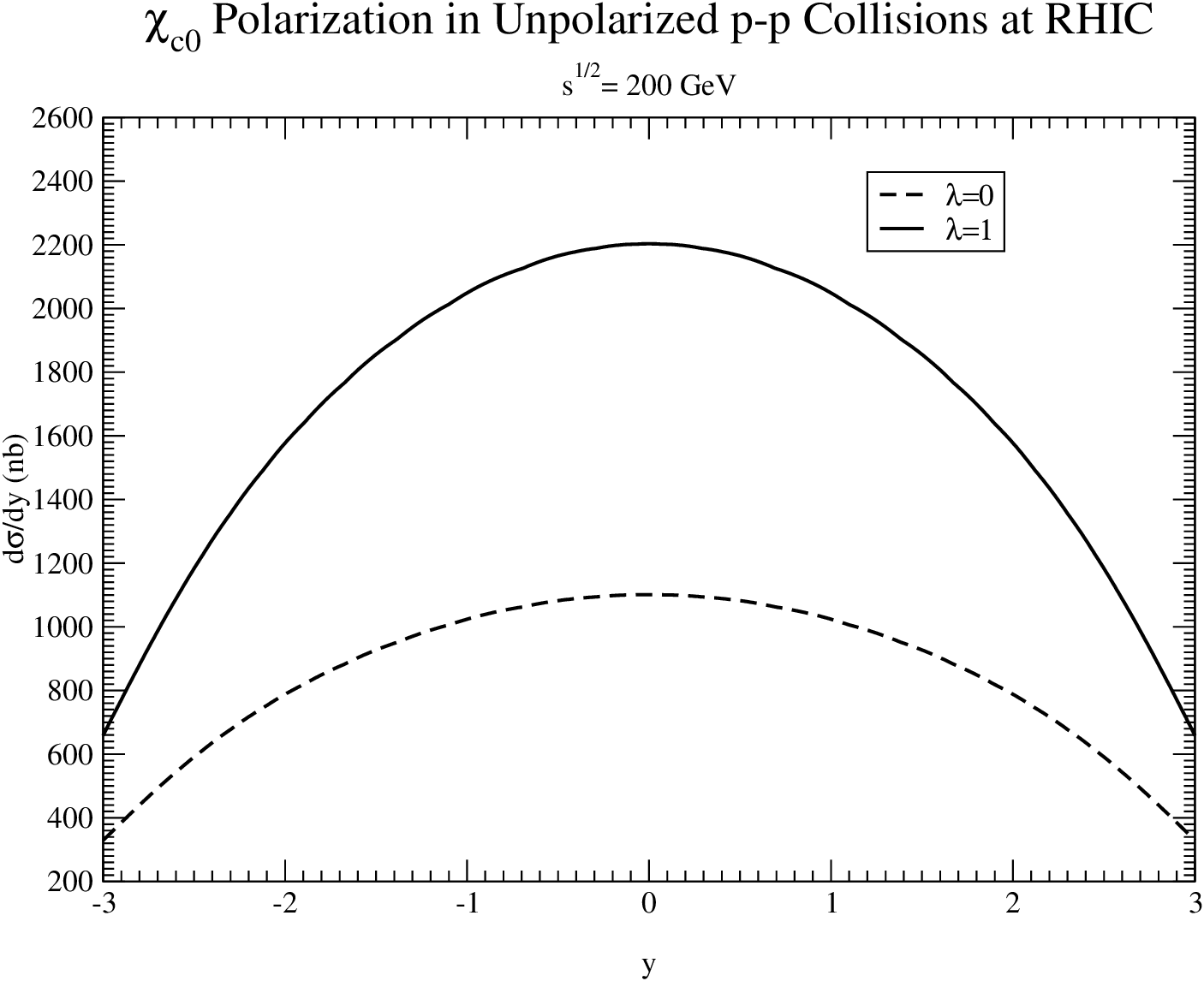,height=10cm}}}
\caption{ Rapidity distribution of $\chi_{c0}$ production cross section at RHIC
in pp collisions at $\sqrt{s}$ = 200 GeV.}
\label{fig1}
\end{figure}

where the polarized partonic matrix element squared involves
the helicity combination $(+,+) - (+,-)$ with $+,-$ denoting the
helicities $h_1$, $h_2$ of the incoming partons \cite{jacknayak}.
Then from eqs. (\ref{eq4}) and (\ref{hel}) we find
\bea
\Delta~|M_{q \bar q \rightarrow Q\bar Q}|^2~=~\frac{ g^4}{4m^2}~
{\eta^\prime}^{\dagger} \sigma^i T^a {\xi^\prime}\,
{\xi}^{\dagger} \sigma^j T^a {\eta}\,~[
2m^2 n_i n_j~ -~\delta_{ij} (k_1 \cdot k_2)]\,,
\label{eq9}
\eea
using $(k_2 \cdot L)_i=-(k_1 \cdot L)_i~=~m~n_i$ were $n_i,n_j$ are the
components of unit three-vectors ${\bf n}_1,{\bf n}_2$
which specify the polarizations of the heavy quarks and heavy
antiquarks respectively in the charmonium bound state.
The leading order term in an expansion in $q$ gives
\bea
\Delta~|M_{q \bar q \rightarrow Q\bar Q}|^2~=~\frac{ g^4}{4}~
[ n_i n_j - \delta_{ij}]~
{\eta^\prime}^{\dagger} \sigma^i T^a {\xi^\prime} \, {\xi}^{\dagger}
\sigma^j T^a {\eta}\,
\label{eqqbar0}
\eea
which after averaging over the initial color (by dividing by 9) gives
\bea
\Delta~|M_{q \bar q \rightarrow Q\bar Q}|^2~=~\frac{ 4\pi^2
\alpha_s^2}{9}~
[ n_i n_j - \delta_{ij}]~
{\eta^\prime}^{\dagger} \sigma^i T^a {\xi^\prime}
{\xi}^{\dagger} \sigma^j T^a {\eta}\,.
\label{eqqbar1}
\eea

\begin{figure}[htb]
\vspace{2pt}
%\centering{\rotatebox{270}{\epsfig{figure=higgspm2.ps,height=7cm}}}
\centering{{\epsfig{figure=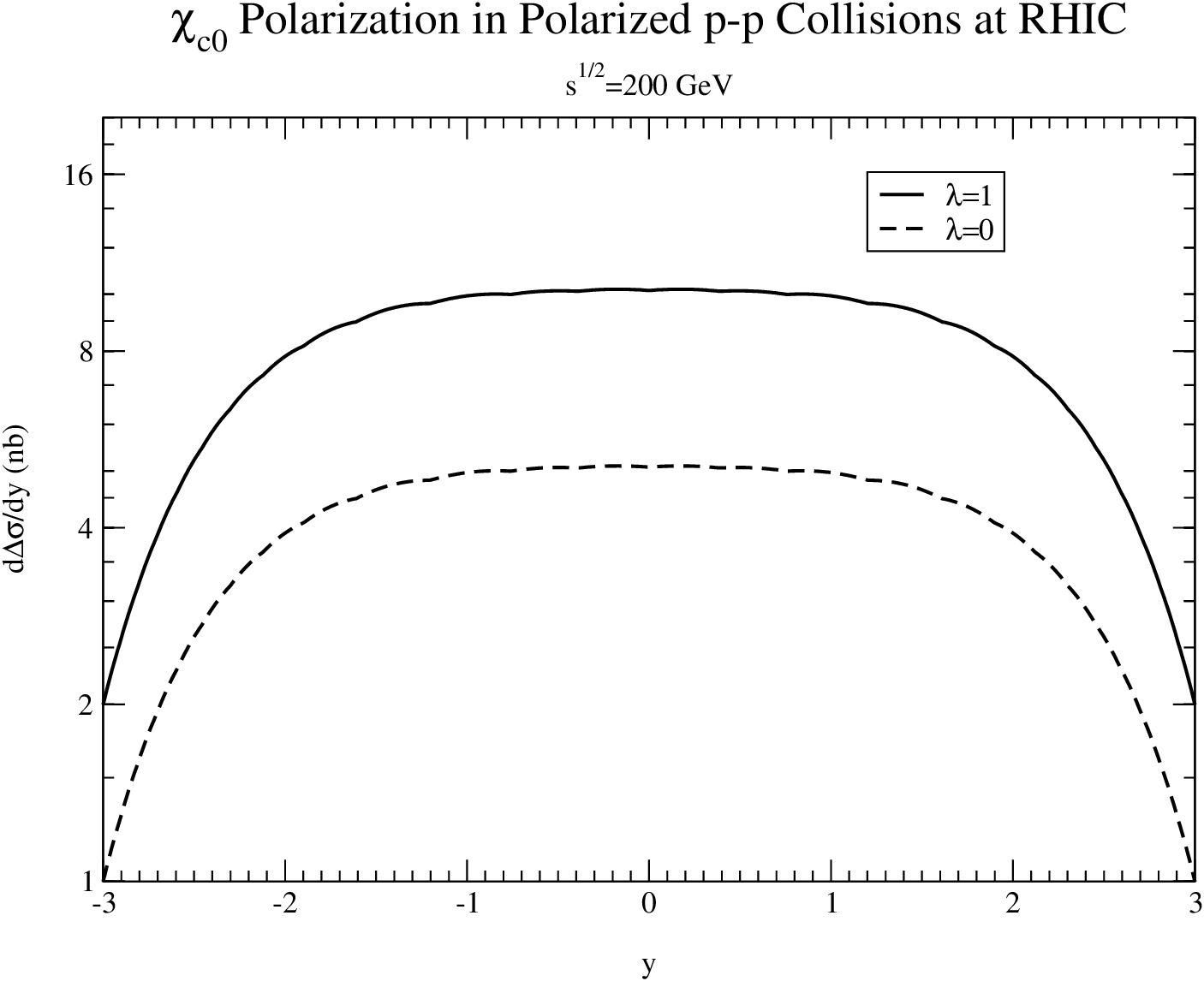,height=10cm}}}
\caption{ Rapidity distribution of $\chi_{c0}$ production cross section at RHIC
in polarized pp collisions at $\sqrt{s}$ = 200 GeV.}
\label{fig2}
\end{figure}

As mentioned in the Appendix B in \cite{braaten}, the two-component spinor factors can be identified with various heavy
quarkonium bound states $H(\lambda)$ with different quantum numbers as
follows
\bea
&&~4m^2~ \eta^{\prime \dagger} \xi^\prime \,\xi^\dagger  \eta ~\equiv~
 < \chi^{\dagger} \psi \,P_{H(\lambda)}\,~\psi^\dagger
 \chi >  ~=~\frac{4}{3} m <{\cal O}^H_1(^1S_0)>\,,
\nonumber \\
&&~4m^2~ \eta^{\prime \dagger} T^a \xi^\prime \,\xi^\dagger T^a \eta ~\equiv~
 < \chi^{\dagger} T^a \psi \,P_{H(\lambda)}\,~\psi^\dagger
T^a \chi >  ~=~\frac{4}{3} m <{\cal O}^H_8(^1S_0)>\,,
\nonumber \\
&&~4m^2~ \eta^{\prime \dagger} \sigma^i \xi^\prime \,\xi^\dagger  \sigma^j
\eta ~\equiv~
 < \chi^{\dagger} \sigma^i \psi \,P_{H(\lambda)}\,~\psi^\dagger
\sigma^{j} \chi >  ~=~\frac{4}{3} U^\dagger_{\lambda i} U_{j\lambda }
m <{\cal O}^H_1(^3S_1)>\,,
\nonumber \\
&&~4m^2~ \eta^{\prime \dagger} \sigma^i T^a \xi^\prime \,\xi^\dagger
\sigma^j T^a \eta ~\equiv~
 < \chi^{\dagger} \sigma^i T^a \psi \,P_{H(\lambda)}\,~\psi^\dagger
\sigma^{j} T^a \chi >  ~=~\frac{4}{3} U^\dagger_{\lambda i} U_{j\lambda }
m <{\cal O}^H_8(^3S_1)>\,,
\nonumber \\
&&~4m^2~q^n {q^m}\eta^{\prime \dagger} \sigma^i \xi^\prime\, \xi^\dagger
\sigma^j
\eta ~\equiv~ < \chi^{\dagger} (-\frac{i}{2}D^m) \sigma^i  \psi
\,P_{H(\lambda)}\,~\psi^\dagger
(-\frac{i}{2}D^n) \sigma^{j}  \chi >  \nonumber \\
&&~=~4 U^\dagger_{\lambda i}\,  U_{j\lambda } \delta^{mn} m
<{\cal O}^H_1(^3P_0)>\,,
\nonumber \\
&&~4m^2~q^n {q^m}\eta^{\prime \dagger} \sigma^i T^a \xi^\prime
\xi^\dagger  \sigma^{j} T^a \eta ~\equiv~ < \chi^{\dagger}
(-\frac{i}{2}D^m) \sigma^i T^a \psi \,
P_{H(\lambda)}~\psi^\dagger  (-\frac{i}{2}D^n) \sigma^{j} T^a \chi >
\nonumber \\
&&~=~4  U^\dagger_{\lambda i} \, U_{j\lambda } \delta^{mn} m
<{\cal O}^H_8(^3P_0)>\,
\label{had}
\eea

\begin{figure}[htb]
\vspace{2pt}
%\centering{\rotatebox{270}{\epsfig{figure=higgspm2.ps,height=7cm}}}
\centering{{\epsfig{figure=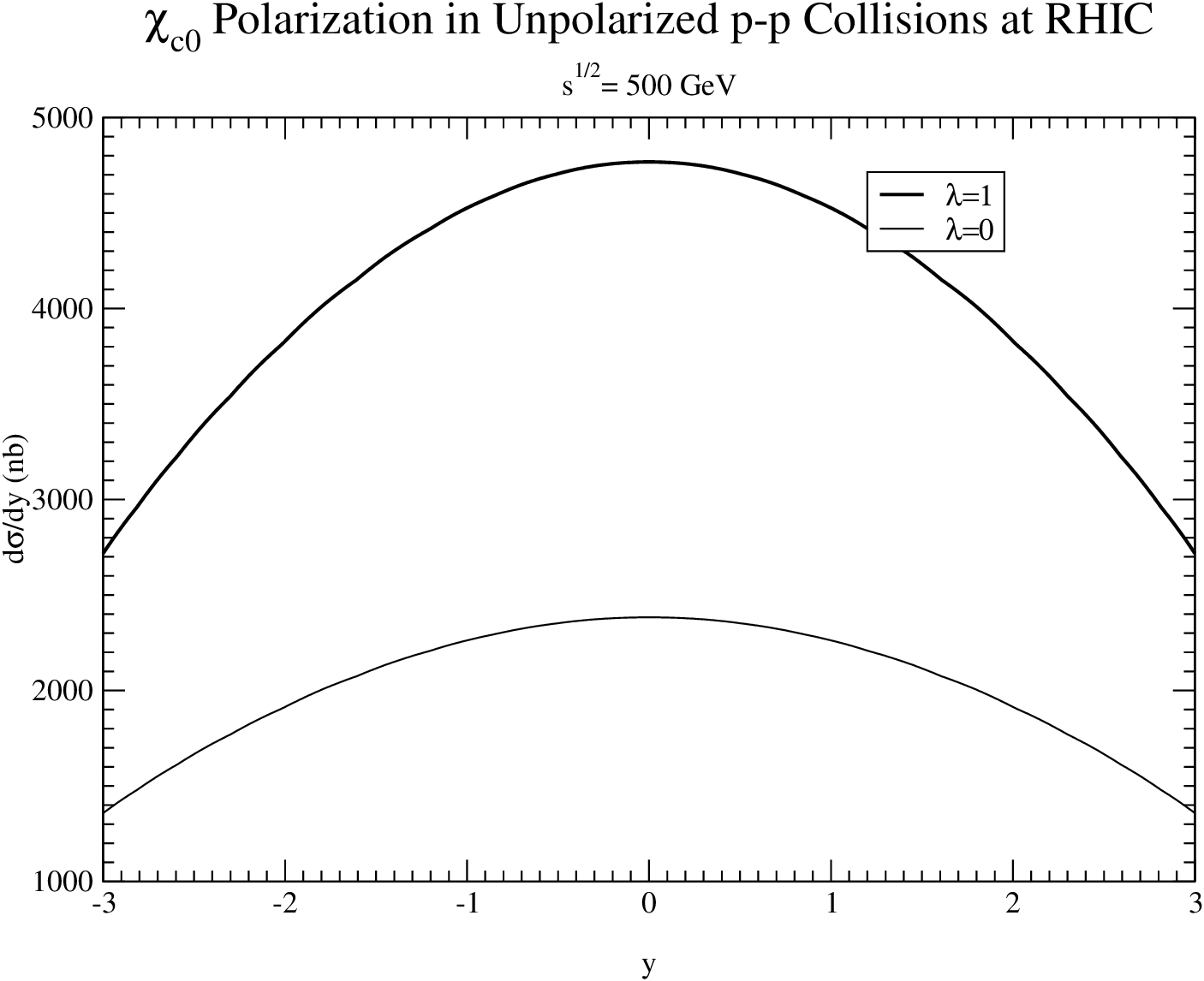,height=10cm}}}
\caption{ Rapidity distribution of $\chi_{c0}$ production cross section at RHIC
in pp collisions at $\sqrt{s}$ = 500 GeV.}
\label{fig3}
\end{figure}

where
\bea
\sum_i U_{\lambda i} U^\dagger_{i \lambda}~=~1
\nonumber \\
\sum_i U_{\lambda i} n^i=\delta_{\lambda 0}\,,
\label{matr}
\eea
where $n^i$ is along the z-direction. Using the above equations
we finally obtain
\bea
\Delta |M_{q \bar q \rightarrow H(\lambda)}|^2~ =~ -\frac{ 4\pi^2
\alpha_s^2}{27}~
[1 - \delta_{\lambda 0}]~<{\cal{O}}^{H}_8(^3S_1)>\,.
\label{eq8}
\eea
The polarized quark-antiquark fusion process cross section for $\chi_{c}(\lambda)$
production is given by
\bea
\Delta \sigma_{q \bar q \rightarrow \chi_{c}(\lambda)}~ =~
-\delta(\hat s -4m^2)~\frac{ \pi^3 \alpha_s^2}{27m^3}~
[1-\delta_{\lambda 0}]~<{\cal{O}}^{\chi_{c}}_8(^3S_1)>\,
\label{eq10}
\eea
which vanishes for $\lambda=0$.

\begin{figure}[htb]
\vspace{2pt}
%\centering{\rotatebox{270}{\epsfig{figure=higgspm2.ps,height=7cm}}}
\centering{{\epsfig{figure=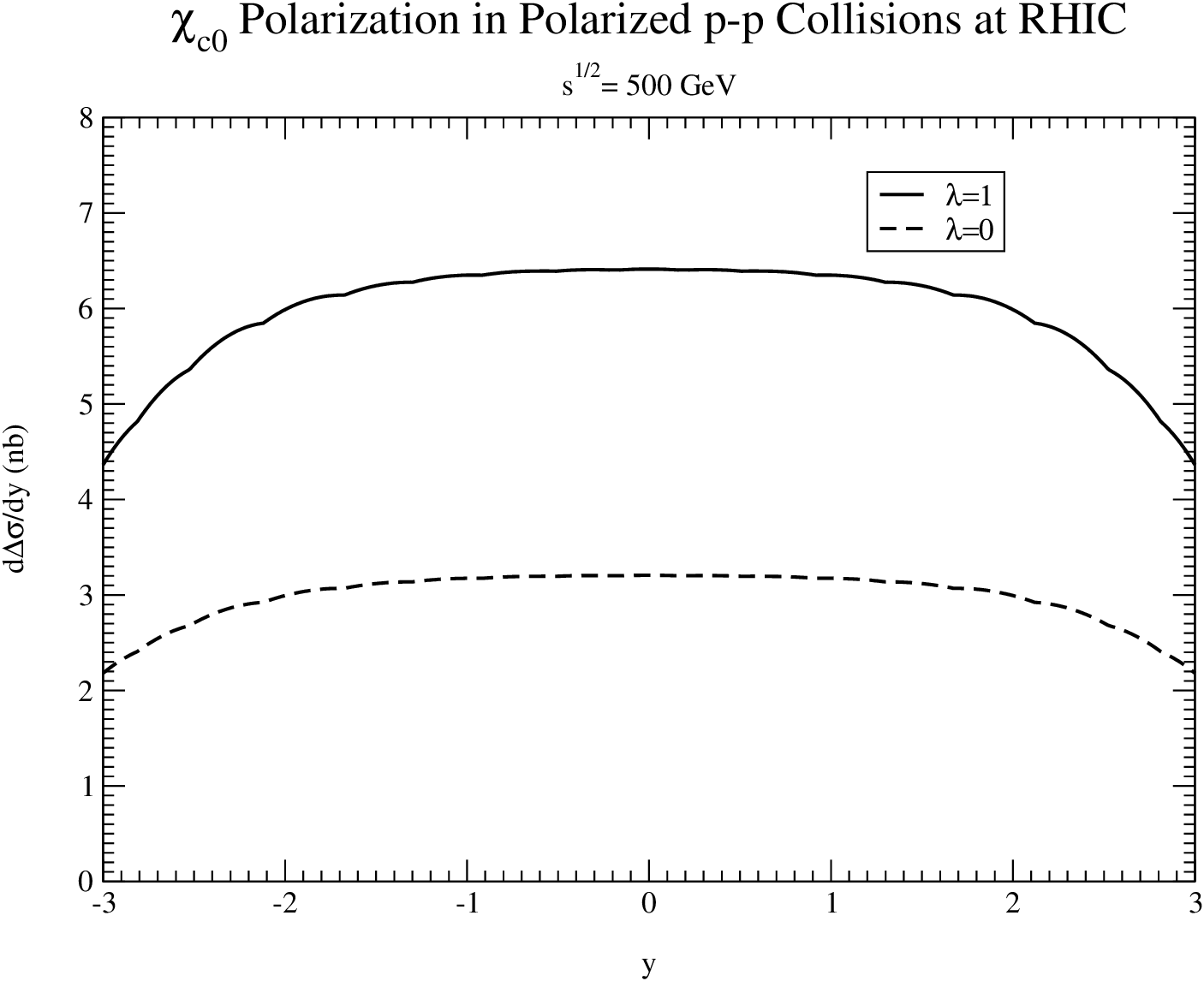,height=10cm}}}
\caption{ Rapidity distribution of $\chi_{c0}$ production cross section at RHIC
in polarized pp collisions at $\sqrt{s}$ = 500 GeV.}
\label{fig4}
\end{figure}

\subsection{The gg fusion process}

At the amplitude level the matrix element for the gluon fusion process $g(k_1) + g(k_2)
\rightarrow Q(p_1) + \bar{Q}(p_2)$
after including s, t, and u channel Feynman diagrams is given by
\bea
M_{gg  \rightarrow Q\bar Q}~=~-g^2~ \epsilon^a_\mu(k_1)
\epsilon^{*b}_\nu(k_2) ~
[(\frac{1}{6}\delta^{ab} ~+~\frac{1}{2}d^{abc}T^c)~S^{\mu \nu}~+~\frac{i}{2}
f^{abc}T^c~F^{\mu \nu}]\,,
\label{eg1p}
\eea
where
\bea
S^{\mu \nu}~=~\bar{u}(p_1)
[\frac{\gamma^\mu ~({{p}\!\!\!\slash}_1 - {{k}\!\!\!\slash}_1 +m)~\gamma^\nu}
{2p_1 \cdot k_1}~+~
\frac{\gamma^\nu ~({{p}\!\!\!\slash}_1 - {{k}\!\!\!\slash}_2 +m)~\gamma^\mu}
{2p_1 \cdot k_2}]\,
v(p_2)
\label{eg2}
\eea
and
\bea
&& F^{\mu \nu}~=~\bar{u}(p_1)
[\frac{\gamma^\mu ~({{p}\!\!\!\slash}_1 - {{k}\!\!\!\slash}_1 +m)~\gamma^\nu}
{2p_1 \cdot k_1}~-~
\frac{\gamma^\nu ~({{p}\!\!\!\slash}_1 - {{k}\!\!\!\slash}_2 +m)~\gamma^\mu}
{2p_1 \cdot k_2}
\nonumber \\
&& ~+~ \frac{2}{P^2}~V^{\mu \nu \lambda}(k_1,k_2,-k_1-k_2)~\gamma_\lambda ]
v(p_2)\,
\label{eg3}
\eea
where the three gluon vertex is denoted by $V^{\mu \nu \lambda}(k_1,k_2,k_3)~=~[
(k_1-k_2)^\lambda g^{\mu \nu}~+~(k_2-k_3)^\mu g^{\nu \lambda}~+~(k_3-k_1)^\nu
g^{\lambda\mu} ]$.

\begin{figure}[htb]
\vspace{2pt}
%\centering{\rotatebox{270}{\epsfig{figure=higgspm2.ps,height=7cm}}}
\centering{{\epsfig{figure=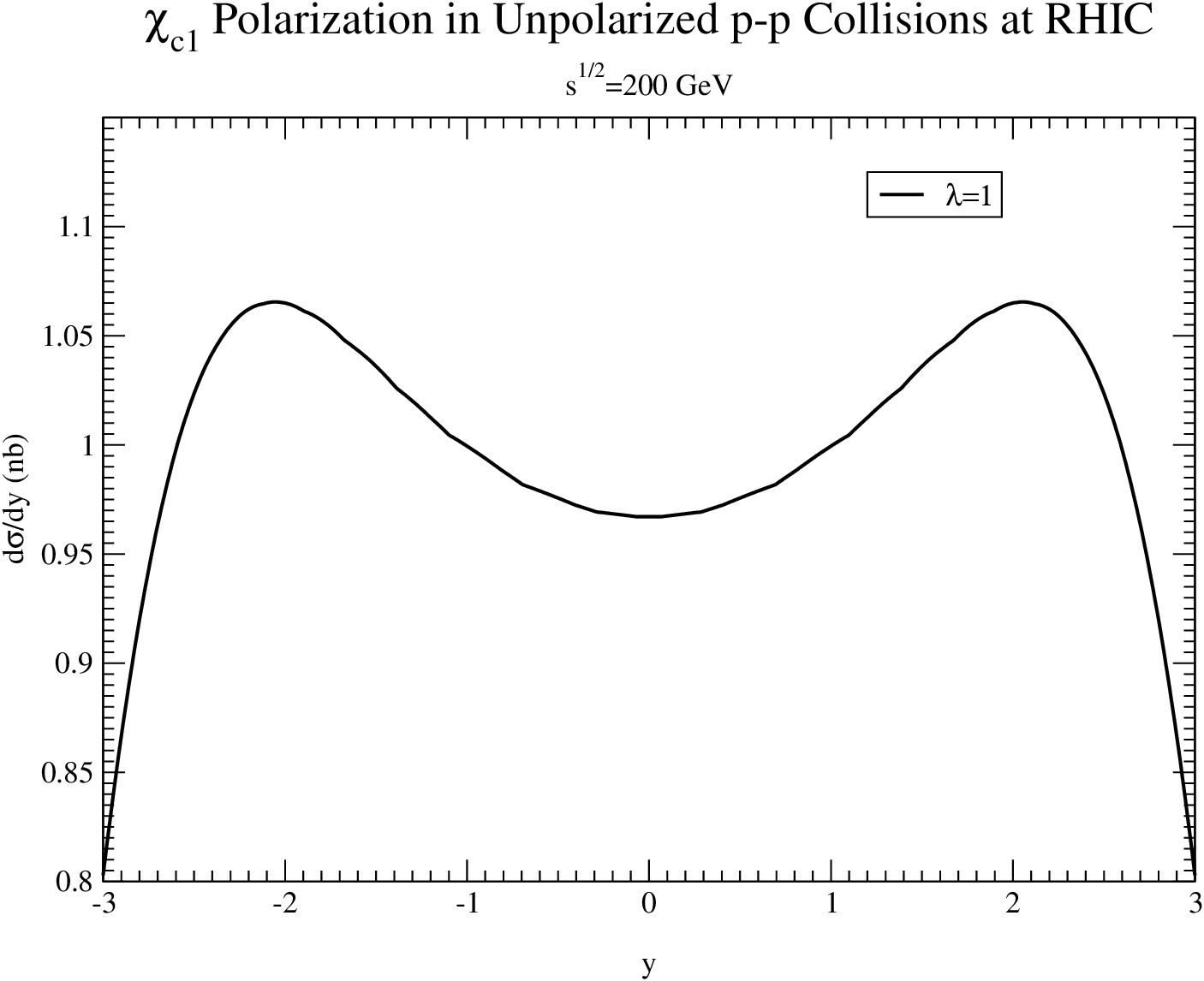,height=10cm}}}
\caption{ Rapidity distribution of $\chi_{c1}$ production cross section at RHIC
in pp collisions at $\sqrt{s}$ = 200 GeV.}
\label{fig5}
\end{figure}

From the identities among the spinors and boost matrices
from the appendix A of \cite{braaten} we find
\bea
&&\bar{u}(p_1) [\frac{\gamma^\mu ~{{k}\!\!\!\slash}_1~\gamma^\nu}
{2p_1 \cdot k_1}~+~
\frac{\gamma^\nu {{k}\!\!\!\slash}_2 ~\gamma^\mu}{2p_1 \cdot k_2}] v(p_2)~=~
\frac{i}{2m^2}(k_1-k_2)_\lambda \epsilon^{\rho \mu \nu \lambda} P_\rho
\xi^\dagger \eta
\nonumber \\
&& ~+~ \frac{(L\cdot k_1)_n}{m^3}~[P^\nu L^\mu_j-P^\mu L^\nu_j
+2g^{\mu \nu}(L\cdot k_1)_j-(k_1-k_2)^\mu
L^\nu_j-(k_1-k_2)^\nu L^\mu_j]~q^n \xi^\dagger \sigma^j \eta
\nonumber \\
&& ~+~ \frac{(L\cdot k_1)_j}{m^3}~[P^\mu L^\nu_n-P^\nu L^\mu_n] ~q^n
\xi^\dagger \sigma^j \eta
~+~ \frac{1}{m}~[P^\mu L^\nu_j-P^\nu L^\mu_j] ~\xi^\dagger \sigma^j \eta\,,
\label{eg4}
\eea
and
\bea
&&\bar{u}(p_1) [\frac{\gamma^\mu ~{{k}\!\!\!\slash}_1~\gamma^\nu}{2p_1
\cdot k_1}~-~
\frac{\gamma^\nu {{k}\!\!\!\slash}_2 ~\gamma^\mu}{2p_1 \cdot k_2}] v(p_2)~=~
\frac{(L \cdot k_1)_n}{2m^4}(k_1-k_2)_\lambda
\epsilon^{\rho \mu \nu \lambda} P_\rho q^n \xi^\dagger \eta
\nonumber \\
&& ~-~ \frac{(L\cdot k_1)_n}{m^3}~[P^\nu L^\mu_j+P^\mu L^\nu_j]~q^n
\xi^\dagger \sigma^j \eta
\nonumber\\
&& -\frac{1}{m}~[2g^{\mu \nu}(L\cdot k_1)_j-(k_1-k_2)^\mu
L^\nu_j-(k_1-k_2)^\nu L^\mu_j]~\xi^\dagger \sigma^j \eta
\nonumber \\
&&~+~ \frac{2}{m}~[L^\mu_n L^\nu_j-L^\nu_n L^\mu_j] ~q^n \xi^\dagger
\sigma^j \eta\,.
\label{eg5}
\eea

\begin{figure}[htb]
\vspace{2pt}
%\centering{\rotatebox{270}{\epsfig{figure=higgspm2.ps,height=7cm}}}
\centering{{\epsfig{figure=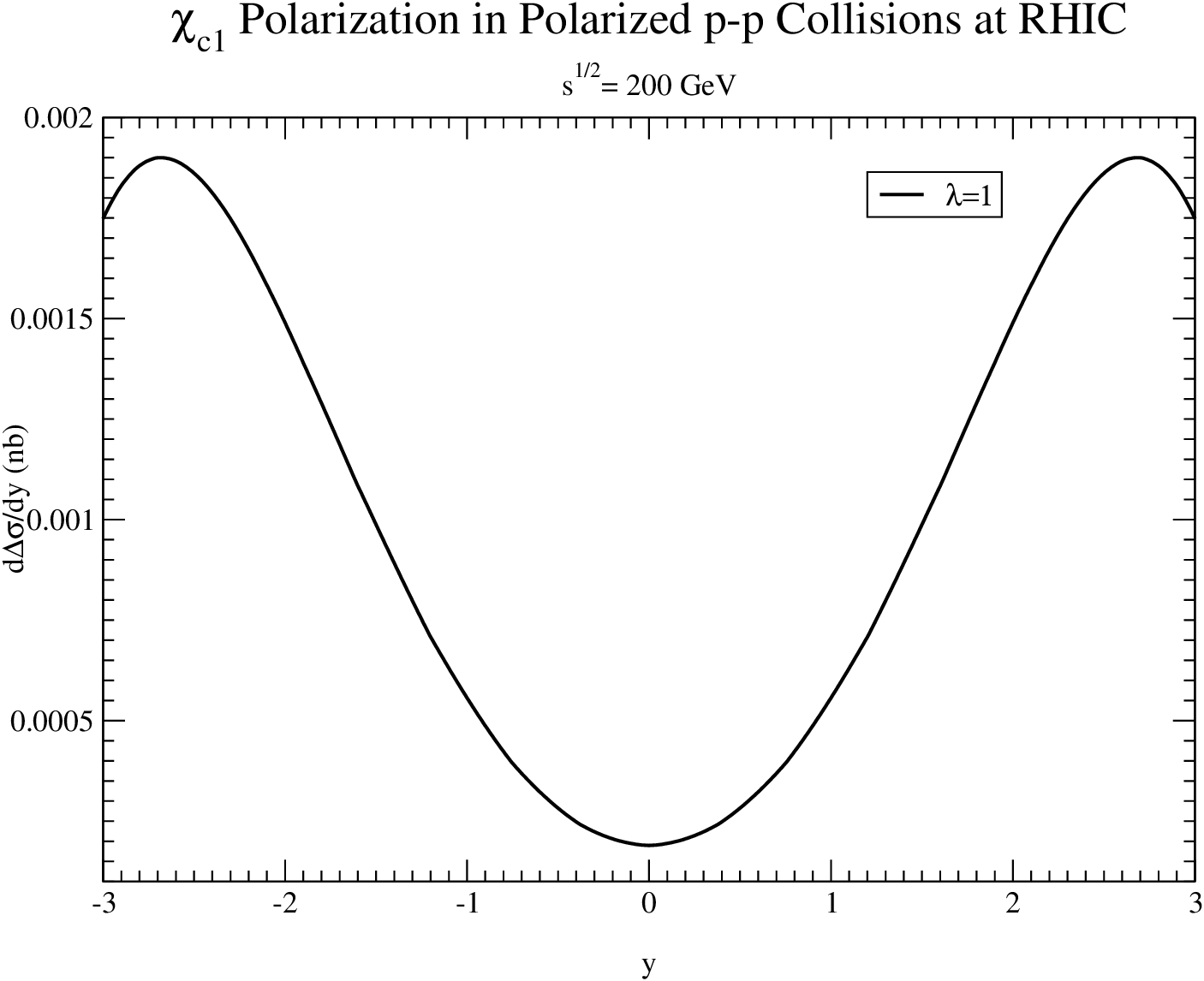,height=10cm}}}
\caption{ Rapidity distribution of $\chi_{c1}$ production cross section at RHIC
in polarized pp collisions at $\sqrt{s}$ = 200 GeV.}
\label{fig6}
\end{figure}

Using above equations we find
\bea
&& S^{\mu \nu}~=~
\frac{i}{2m^2}(k_1-k_2)_\lambda \epsilon^{\rho \mu \nu \lambda} P_\rho
\xi^\dagger \eta
~+~ [\frac{(L\cdot k_1)_j}{m^3}~(P^\nu L^\mu_n-P^\mu L^\nu_n-2g^{\mu \nu}
(L\cdot k_1)_n)
\nonumber \\
&& ~+~ \frac{2}{m}~[L^\mu_n L^\nu_j+L^\nu_n L^\mu_j]
+\frac{1}{m^3}(L \cdot k_1)_n[(k_1-k_2)^\mu L^\nu_j+(k_1-k_2)^\nu L^\mu_j]]~q^n
\xi^\dagger \sigma^j \eta\,,
\label{eg6}
\eea
and
\bea
F^{\mu \nu}~=~
\frac{i(L \cdot k_1)_n}{2m^4}(k_1-k_2)_\lambda \epsilon^{\rho \mu \nu \lambda}
P_\rho q^n \xi^\dagger \eta
+[k_2^\nu L^\mu_j-k_1^\mu L^\nu_j]~\xi^\dagger \sigma^j \eta \,.
\label{eg7}
\eea
The square of gluon polarization vector, for an incoming gluon with a helicity $\lambda_i$,
can be written as \cite{sivers}
\bea
\epsilon^a_\mu(k_1,\lambda_i)
\epsilon^{*b}_\nu(k_1,\lambda_i)
~=~\frac{1}{2}~\delta^{ab}~
[-g_{\mu \nu}~+\frac{k_{1 \mu} k_{2\nu}+k_{2\mu}k_{1\nu}}{k_1 \cdot k_2}-
i\lambda_i ~\epsilon_{\mu \nu \rho \delta}
\frac{k_1^\rho k_2^\delta}{k_1 \cdot k_2}]\,.
\label{nopolsump}
\eea

\begin{figure}[htb]
\vspace{2pt}
%\centering{\rotatebox{270}{\epsfig{figure=higgspm2.ps,height=7cm}}}
\centering{{\epsfig{figure=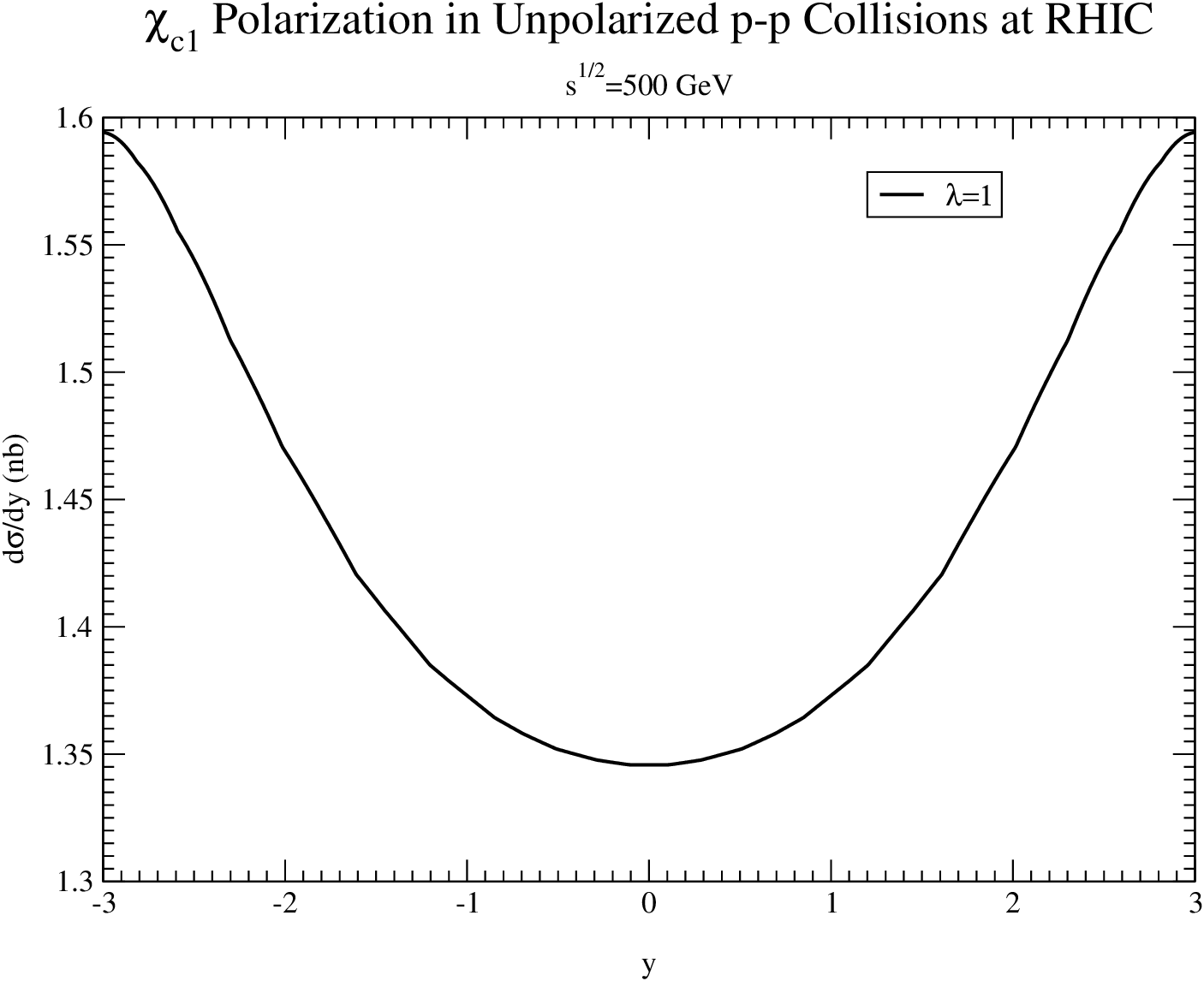,height=10cm}}}
\caption{ Rapidity distribution of $\chi_{c1}$ production cross section at RHIC
in pp collisions at $\sqrt{s}$ = 500 GeV.}
\label{fig7}
\end{figure}

Choosing longitudinally polarized gluons and using the relation
\bea
\epsilon_{\mu \mu^\prime \alpha \beta} ~k_1^\alpha k_2^\beta~=~2m^2
\epsilon^{ijk} n_k L^\mu_i L^\nu_j\,,
\label{eps2}
\eea
from appendix A of \cite{braaten} we find that
\bea
&&~ \Delta |M_{gg  \rightarrow Q\bar Q}|^2~=~-\frac{g^4}{4}~
\epsilon^{pqr} \epsilon^{p^\prime q^\prime r^\prime} n_r n_{r^\prime}
\nonumber \\
&&~\times[S^{ab}S^{*ab} L_{\mu p} L_{\nu p^\prime} S^{\mu \nu} L_{\mu^\prime q}
L_{\nu^\prime q^\prime}
S^{* \mu^\prime \nu^\prime} +F^{ab}F^{*ab} L_{\mu p} L_{\nu p^\prime}
F^{ \mu \nu}
L_{\mu^\prime q} L_{\nu^\prime q^\prime} F^{*\mu^\prime \nu^\prime}]\,,
\label{dm3}
\eea
where
\bea
S^{ab}~=~\frac{1}{6}\delta^{ab}~+~\frac{1}{2}d^{abc}T^c
~~~~~~~{\rm and }~~~~~~~~~ F^{ab}~=~\frac{i}{2}f^{abc}T^c\,.
\label{sf}
\eea

\begin{figure}[htb]
\vspace{2pt}
%\centering{\rotatebox{270}{\epsfig{figure=higgspm2.ps,height=7cm}}}
\centering{{\epsfig{figure=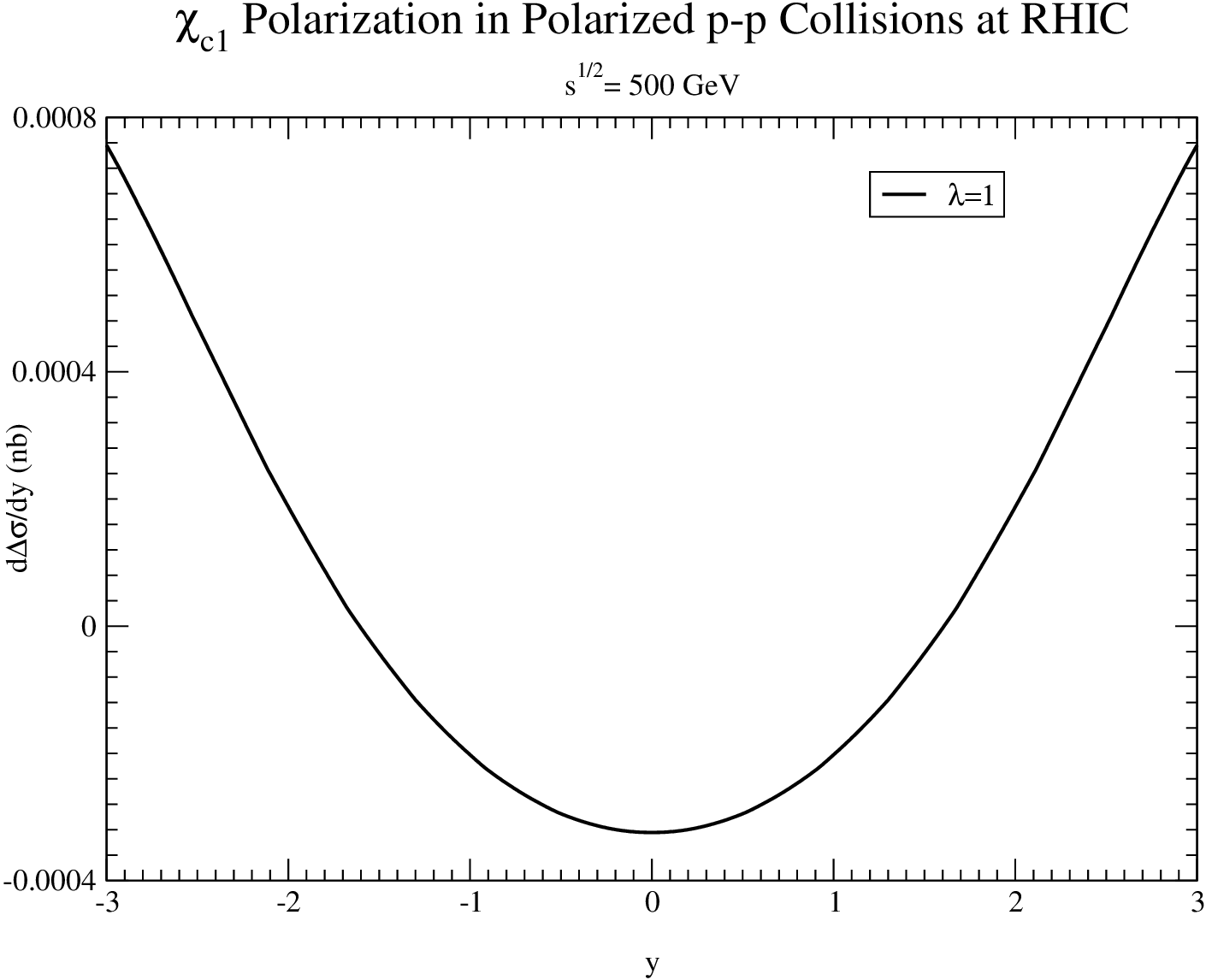,height=10cm}}}
\caption{ Rapidity distribution of $\chi_{c1}$ production cross section at RHIC
in polarized pp collisions at $\sqrt{s}$ = 500 GeV.}
\label{fig8}
\end{figure}

Using the properties of $L^\mu_i$ matrices \cite{braaten}
and after averaging over the initial color (by dividing by 64)
we find
\bea
&& \Delta |M_{gg  \rightarrow Q\bar Q}|^2~=~-\frac{\pi^2 \alpha_s^2}{9}~[
\eta^{\prime \dagger} \xi^\prime \xi^\dagger \eta +
\frac{1}{m^2}[(n \cdot q) n_jq^\prime_{j^\prime}
 +(n \cdot q^\prime)n_{j^\prime} q_j -\frac{3}{2}(n \cdot q)
(n \cdot q^\prime)n_j n_{j^\prime}
\nonumber \\
&& -(n \times q^\prime)_j (n \times q)_{j^\prime}]
\eta^{\prime \dagger} \sigma^{j^\prime} \xi^\prime \xi^\dagger
\sigma^{j} \eta +\frac{15}{8}
\eta^{\prime \dagger} T^a \xi^\prime \xi^\dagger T^a \eta +\frac{15}{8m^2}
[(n \cdot q) n_jq^\prime_{j^\prime}
 +(n \cdot q^\prime)n_{j^\prime} q_j
\nonumber \\
&& -\frac{3}{2}(n \cdot q) (n \cdot q^\prime)n_j n_{j^\prime}
 -(n \times q^\prime)_j (n \times q)_{j^\prime}]
\eta^{\prime \dagger} \sigma^{j^\prime} T^a \xi^\prime \xi^\dagger
\sigma^{j} T^a \eta
\nonumber \\
&& +\frac{27}{8m^2}~
(n \cdot q) (n \cdot q^\prime)
\eta^{\prime \dagger} T^a \xi^\prime \xi^\dagger  T^a \eta ]\,.
\label{dpol}
\eea

While for $j/\psi$ and $\psi'$ production the matrix elements $<{\cal{O}}^{j/\psi (\psi')}_8(^1S_0)>$,
$<{\cal{O}}^{j/\psi (\psi')}_8(^3S_1)>$ and $<{\cal{O}}^{j/\psi (\psi')}_8(^3P_0)>$
are important \cite{jacknayak}, for $\chi_{c0}$ production the matrix elements
$<{\cal{O}}^{\chi_{c0}}_8(^3S_1)>$ and $<{\cal{O}}^{\chi_{c0}}_1(^3P_0)>$ are important
\cite{jungil}.

\begin{figure}[htb]
\vspace{2pt}
%\centering{\rotatebox{270}{\epsfig{figure=higgspm2.ps,height=7cm}}}
\centering{{\epsfig{figure=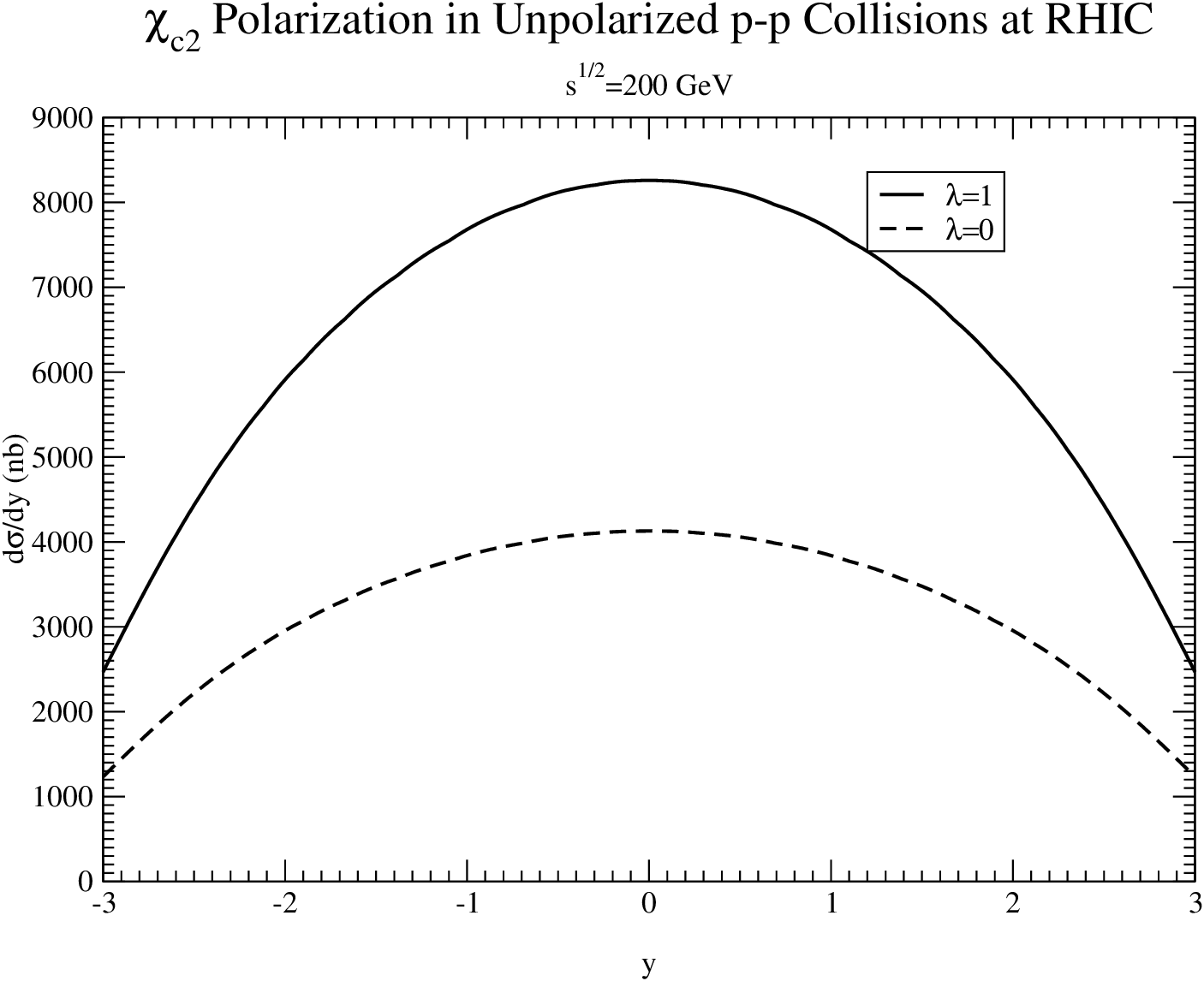,height=10cm}}}
\caption{ Rapidity distribution of $\chi_{c2}$ production cross section at RHIC
in pp collisions at $\sqrt{s}$ = 200 GeV.}
\label{fig9}
\end{figure}

We identify different bound states as given in eq. (\ref{had})
and use eq. (\ref{matr}) to obtain
\bea
&& \Delta \sigma_{gg  \rightarrow \chi_0(\lambda)}~=~-\delta(\hat s -4m^2)
~\frac{\pi^3 \alpha_s^2}{36m^5}~(\frac{1}{2} \delta_{\lambda 0}-1) <{\cal O}^{\chi_{c0}}_1(^3P_0)>\,.
\label{dplf0}
\eea
and
\bea
&& \Delta \sigma_{gg  \rightarrow \chi_2(\lambda)}~=~-\delta(\hat s -4m^2)
~\frac{\pi^3 \alpha_s^2}{135m^5}~(\frac{1}{2} \delta_{\lambda 0}-1) <{\cal O}^{\chi_{c2}}_1(^3P_2)>\,.
\label{dplf2}
\eea

\subsection{The Polarized Proton-Proton Collisions}

Folding eqs. (\ref{eq10}), (\ref{dplf0}) with parton densities
we find the following cross section for $\chi_{c0}(\lambda)$ in
longitudinally polarized proton-proton collisions
\bea
&& \Delta \sigma_{(pp \rightarrow \chi_{c0}(\lambda))}~=~\frac{\pi^3 \alpha_s^2}{27sm^3}~
\int_{4m^2/s}^1 \frac{dx_1 }{x_1}
~[\Delta f_q(x_1,2m) ~\Delta f_{\bar q}(\frac{4m^2}{x_1 s},2m)
\nonumber \\
&& ~\times~ (\delta_{\lambda 0}-1)\,<{\cal{O}}^{\chi_{c0}}_8(^3S_1)>
~+~\frac{1}{4}~\Delta f_g(x_1,2m) ~\Delta f_{g}(\frac{4m^2}{x_1 s},2m)
\nonumber \\
&&~\times~ \frac{9}{m^2} (1-\frac{1}{2} \delta_{\lambda 0})
<{\cal O}^{\chi_{c0}}_1(^3P_0)>]\,,
\label{polfin}
\eea
where $\Delta f (x,Q)(\Delta g(x,Q))$ denote the polarized quark (gluon)
distribution functions inside the proton at the scale $Q$.

\begin{figure}[htb]
\vspace{2pt}
%\centering{\rotatebox{270}{\epsfig{figure=higgspm2.ps,height=7cm}}}
\centering{{\epsfig{figure=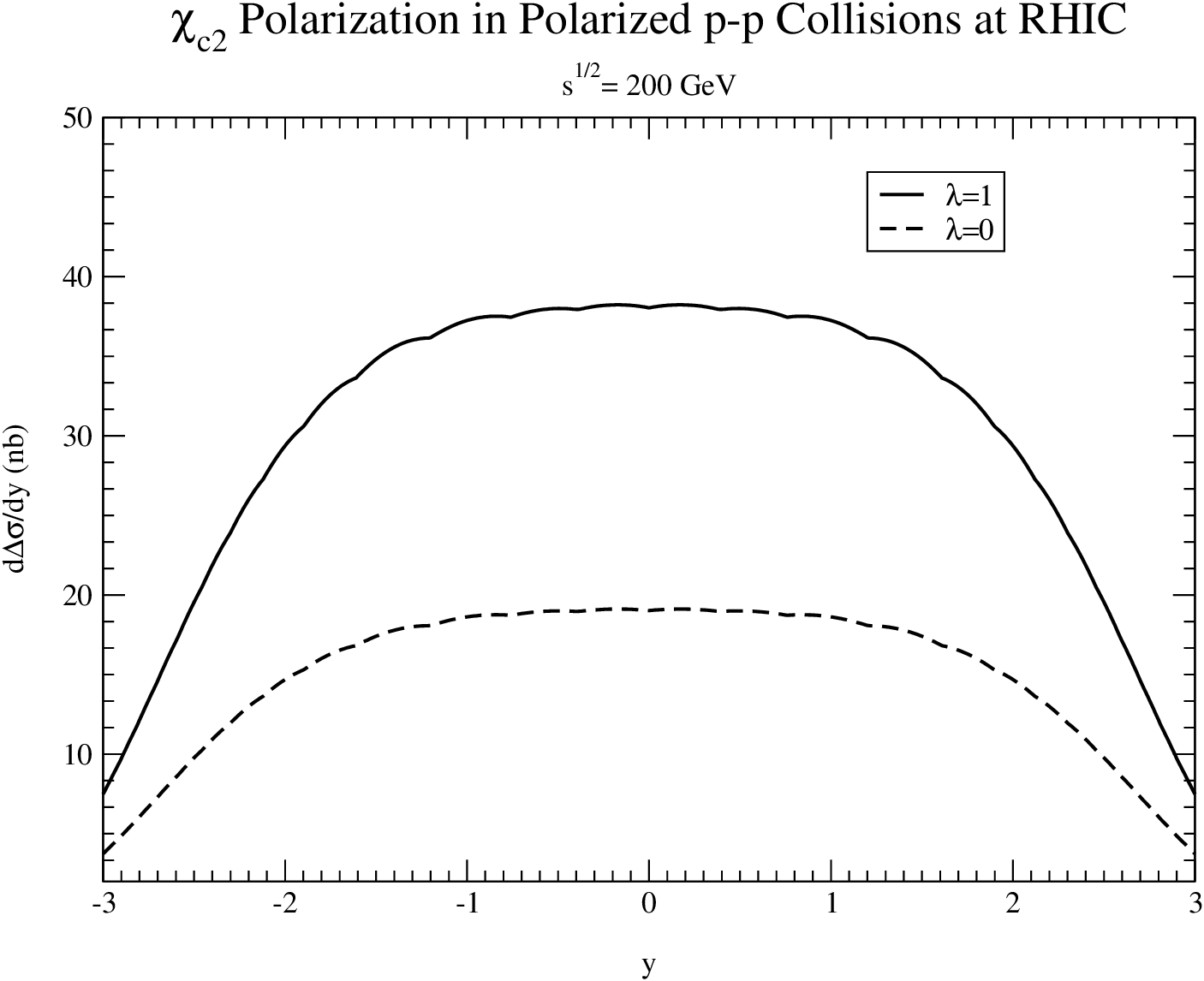,height=10cm}}}
\caption{ Rapidity distribution of $\chi_{c2}$ production cross section at RHIC
in polarized pp collisions at $\sqrt{s}$ = 200 GeV.}
\label{fig10}
\end{figure}

Folding eqs. (\ref{eq10}) with parton densities
we find the following cross section for $\chi_{c1}(\lambda)$ in
longitudinally polarized proton-proton collisions
\bea
&& \Delta \sigma_{(pp \rightarrow \chi_{c1}(\lambda))}~=~\frac{\pi^3 \alpha_s^2}{27sm^3}~
\int_{4m^2/s}^1 \frac{dx_1 }{x_1}
~[\Delta f_q(x_1,2m) ~\Delta f_{\bar q}(\frac{4m^2}{x_1 s},2m)
\nonumber \\
&& ~\times~ (\delta_{\lambda 0}-1)\,<{\cal{O}}^{\chi_{c1}}_8(^3S_1)>]\,,
\label{polfin1}
\eea

Similarly, folding eqs. (\ref{eq10}), (\ref{dplf2}) with parton densities
we find the following cross section for $\chi_{c2}(\lambda)$ in
longitudinally polarized proton-proton collisions
\bea
&& \Delta \sigma_{(pp \rightarrow \chi_{c2}(\lambda))}~=~\frac{\pi^3 \alpha_s^2}{27sm^3}~
\int_{4m^2/s}^1 \frac{dx_1 }{x_1}
~[\Delta f_q(x_1,2m) ~\Delta f_{\bar q}(\frac{4m^2}{x_1 s},2m)
\nonumber \\
&& ~\times~ (\delta_{\lambda 0}-1)\,<{\cal{O}}^{\chi_{c2}}_8(^3S_1)>
~+~\frac{1}{15}~\Delta f_g(x_1,2m) ~\Delta f_{g}(\frac{4m^2}{x_1 s},2m)
\nonumber \\
&&~\times~ \frac{9}{m^2} (1-\frac{1}{2} \delta_{\lambda 0})
<{\cal O}^{\chi_{c2}}_1(^3P_2)>]\,,
\label{polfin2}
\eea

The corresponding production cross sections
for unpolarized proton-proton collisions are \cite{braaten}:
\bea
&& \sigma_{(pp \rightarrow \chi_0(\lambda))}~=~\frac{\pi^3 \alpha_s^2}{27sm^3}~
\int_{4m^2/s}^1 \frac{dx_1 }{x_1}
~[f_q(x_1,2m) ~f_{\bar q}(\frac{4m^2}{x_1 s},2m)
\nonumber \\
&&~\times~(1-\delta_{\lambda 0})~<{\cal{O}}^{\chi_{c0}}_8(^3S_1)>
~+~\frac{1}{4}~f_g(x_1,2m) ~ f_{g}(\frac{4m^2}{x_1 s},2m)
\nonumber \\
&&~\times~ \frac{9}{m^2} (1-\frac{2}{3} \delta_{\lambda 0})
<{\cal O}^{\chi_{c0}}_1(^3P_0)> ]\,,
\label{unpolfin}
\eea

\bea
&& \sigma_{(pp \rightarrow \chi_1(\lambda))}~=~\frac{\pi^3 \alpha_s^2}{27sm^3}~
\int_{4m^2/s}^1 \frac{dx_1 }{x_1}
~[f_q(x_1,2m) ~f_{\bar q}(\frac{4m^2}{x_1 s},2m)
\nonumber \\
&&~\times~(1-\delta_{\lambda 0})~<{\cal{O}}^{\chi_{c1}}_8(^3S_1)>]\,
\label{unpolfin1}
\eea
and
\bea
&& \sigma_{(pp \rightarrow \chi_2(\lambda))}~=~\frac{\pi^3 \alpha_s^2}{27sm^3}~
\int_{4m^2/s}^1 \frac{dx_1 }{x_1}
~[f_q(x_1,2m) ~f_{\bar q}(\frac{4m^2}{x_1 s},2m)
\nonumber \\
&&~\times~(1-\delta_{\lambda 0})~<{\cal{O}}^{\chi_{c2}}_8(^3S_1)>
~+~\frac{1}{15}~f_g(x_1,2m) ~ f_{g}(\frac{4m^2}{x_1 s},2m)
\nonumber \\
&&~\times~ \frac{9}{m^2} (1-\frac{2}{3} \delta_{\lambda 0})
<{\cal O}^{\chi_{c2}}_1(^3P_0)> ]\,.
\label{unpolfin2}
\eea

\begin{figure}[htb]
\vspace{2pt}
%\centering{\rotatebox{270}{\epsfig{figure=higgspm2.ps,height=7cm}}}
\centering{{\epsfig{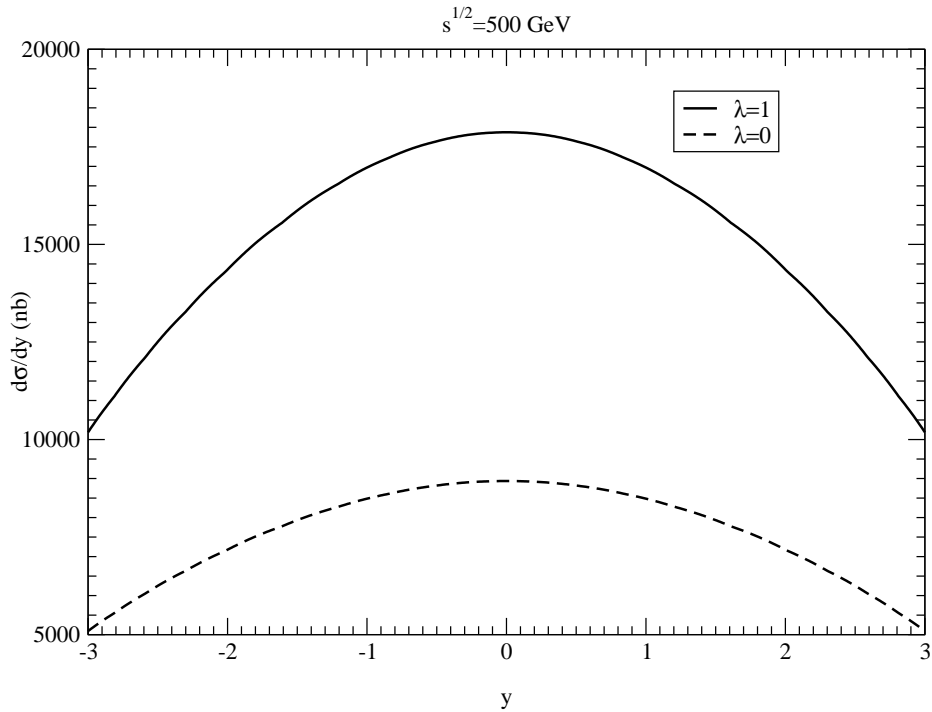}}}
\caption{ Rapidity distribution of $\chi_{c2}$ production cross section at RHIC
in pp collisions at $\sqrt{s}$ = 500 GeV.}
\label{fig11}
\end{figure}

The spin asymmetry $A_{LL}(\lambda)$ is given by the ratio of the
above cross sections
\bea
A_{LL}(\lambda)~=~\frac{d \Delta \sigma(\lambda)}{d \sigma(\lambda)}\,.
\label{spinasym}
\eea

\section{Results and Discussion}

In this section, using the formulae derived above, we compute the LO rapidity
distributions and spin asymmetries for the heavy charmonium systems
$\chi_{cJ}$ in longitudinally polarized proton-proton
collisions at RHIC. We present the results for the $\chi_{c0}$, $\chi_{c1}$ and $\chi_{c2}$
production with definite polarizations in unpolarized as well as polarized proton-proton
collisions at $\sqrt{s}$ = 200 GeV and 500 GeV at RHIC.
These results provide interesting information on the
polarization state of these heavy charmonium states.

We use the following values for the NRQCD non-perturbative matrix elements.
From the Fermilab Tevatron, see \cite{BK} and \cite{CL},
the central values for $\chi_{cJ}$ production the non-perturbative
matrix elements are given by \cite{kniehl}
\bea
<{\cal{O}}^{\chi_{c0}}_8(^3S_1)>=0.0019 ~~GeV^3,~~~~~~~~~~~~~~~~~
<{\cal{O}}^{\chi_{c0}}_1(^3P_0)>=0.089 ~~GeV^5.
\label{mv}
\eea
The non-perturbative matrix elements for $\chi_{c1}$ and $\chi_{c2}$ can be obtained by using symmetry as
follows
\bea
<{\cal{O}}^{\chi_{cJ}}_8(^3S_1)>=(2J+1) <{\cal{O}}^{\chi_{c0}}_8(^3S_1)>,~~~~~~~~~<{\cal{O}}^{\chi_{cJ}}_1(^3P_J)>=(2J+1) <{\cal{O}}^{\chi_{c0}}_1(^3P_0)>. \nonumber \\
\label{mv1}
\eea

\subsection{Rapidity Distribution of $\chi_{cJ}$ Cross Sections at RHIC }

We will present our results in the rapidity range $-3 <~y~<3$.
This covers the central arm (forward arm) electron (muon) detector at the PHENIX
experiment for the $\chi_{cJ}$ rapidity range $-0.5<~y~<0.5$ ($1<~|y|~<2$).
We will present our differential rapidity distributions and spin asymmetries
for $\chi_{cJ}$ production with
helicities $\lambda$ = 1 and 0 in unpolarized and polarized p-p
collisions at $\sqrt s$ = 200 GeV and 500 GeV in the above detector
acceptance ranges.

\begin{figure}[htb]
\vspace{2pt}
%\centering{\rotatebox{270}{\epsfig{figure=higgspm2.ps,height=7cm}}}
\centering{{\epsfig{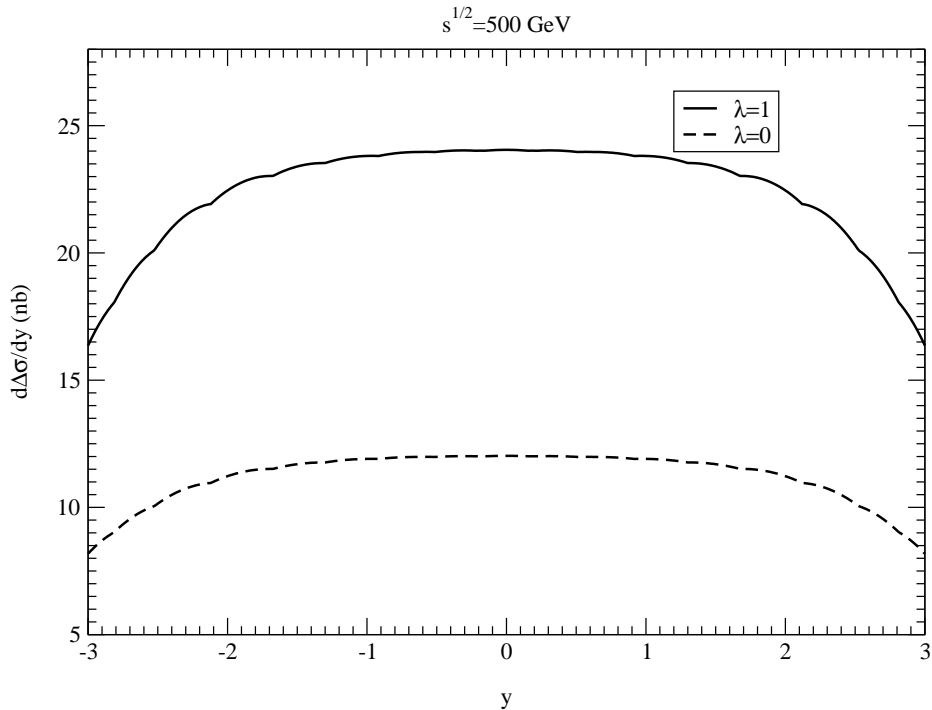}}}
\caption{ Rapidity distribution of $\chi_{c2}$ production cross section at RHIC
in polarized pp collisions at $\sqrt{s}$ = 500 GeV.}
\label{fig12}
\end{figure}

We take the charm quark mass $m$=1.5 GeV and the mass factorization scale
equal to $2 m$. Several groups have produced polarized parton density sets
\cite{GS},\cite{grsv} and \cite{blbo}.  We choose the GRV unpolarized LO
parton densities \cite{grv98} and the GRSV \cite{grsv} polarized densities.
The latter authors have a standard scenario and a valence scenario.
For simplicity we choose the former. Therefore we always use the LO four
flavour sets (for the u, d, s and g partons)
and we set $n_f=4$ in the one-loop running coupling constant and the
parton densities.  For both parton density sets  we use
$\Lambda^{\rm LO}_4 = 175 $ MeV, so that
$\alpha_s^{\rm LO}(m_Z) = 0.121$ at the mass of the Z.

In Fig. 1 we present the rapidity differential distributions
for $\chi_{c0}$ production in unpolarized p-p collisions
at $\sqrt s$ = 200 GeV. The solid and dashed lines correspond to
$\lambda$=1 and 0 respectively. Note that for $\lambda$=0 the cross
section becomes small because the color octet contribution from the
quark-antiquark process at LO vanishes, see eq. (\ref{unpolfin}). The
color singlet contribution is from gluon fusion process at LO, see
eq. (\ref{unpolfin}).

In Fig. 2 we present the rapidity differential distributions
for $\chi_{c0}$ production in polarized p-p collisions
at $\sqrt s$ = 200 GeV. The solid and dashed lines correspond to
$\lambda$=1 and 0 respectively. Note that for $\lambda$=0 the cross
section becomes small because the color octet contribution from the
quark-antiquark process at LO vanishes, see eq. (\ref{eq10}). The
color singlet contribution is from gluon fusion process at LO, see
eq. (\ref{polfin}).

In Fig. 3 we present the rapidity differential distributions
for $\chi_{c0}$ production in unpolarized p-p collisions
at $\sqrt s$ = 500 GeV. The solid and dashed lines correspond to
$\lambda$=1 and 0 respectively. We find that the cross section
for $\chi_{c0}$ production in unpolarized p-p collisions at $\sqrt{s}$=500
GeV is larger than that at $\sqrt{s}$=200 GeV. This is due to the
enhancement of parton distribution function.

In Fig. 4 we present the
rapidity differential distributions for $\chi_{c0}$ production in polarized p-p collisions
at $\sqrt s$ = 500 GeV. The solid and dashed lines correspond to
$\lambda$=1 and 0 respectively. We find that the cross section
for $\chi_{c0}$ production in polarized p-p collisions at $\sqrt{s}$=500
GeV is smaller than that at $\sqrt{s}$=200 GeV. This is due to the
polarized parton distribution function.

\begin{figure}[htb]
\vspace{2pt}
%\centering{\rotatebox{270}{\epsfig{figure=higgspm2.ps,height=7cm}}}
\centering{{\epsfig{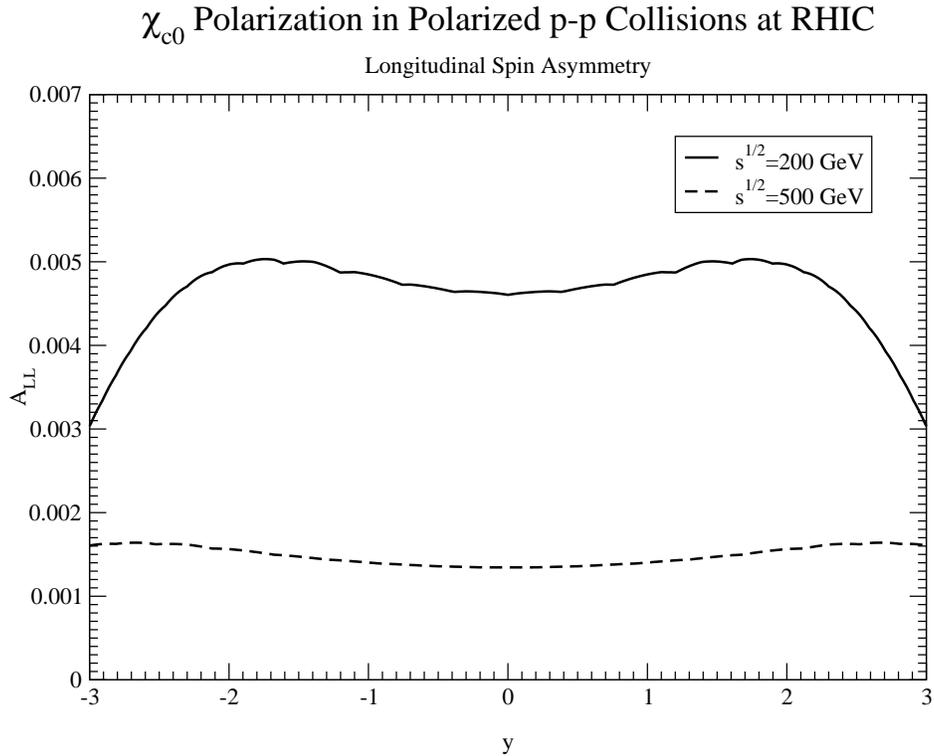}}}
\caption{ Rapidity distribution of longitudinal spin asymmetry $A_{LL}$
of $\chi_{c0}$ production at RHIC in polarized pp collisions.}
\label{fig13}
\end{figure}

In Fig. 5 we present the rapidity differential distributions
for $\chi_{c1}$ production in unpolarized p-p collisions
at $\sqrt s$ = 200 GeV. Note that the shape of the curve is different
than that from $\chi_{c0}$ production because the contribution to $\chi_{c1}$
production is from quark-antiquark fusion process via color octet
mechanism. At LO the gluon fusion process in the color singlet channel
does not contribute to the $\chi_{c1}$ production, see eq. (\ref{unpolfin1}).

In Fig. 6 we present the rapidity differential distributions
for $\chi_{c1}$ production in polarized p-p collisions
at $\sqrt s$ = 200 GeV. In case of $\chi_{c1}$ production the $\lambda$=1
contributes because for $\lambda$=0 the cross section in eq. (\ref{eq10})
in polarized p-p collisions vanishes. At LO the gluon fusion process in the
color singlet channel does not contribute to the $\chi_{c1}$ production in
polarized p-p collisions, see eq. (\ref{polfin1}).

\begin{figure}[htb]
\vspace{2pt}
%\centering{\rotatebox{270}{\epsfig{figure=higgspm2.ps,height=7cm}}}
\centering{{\epsfig{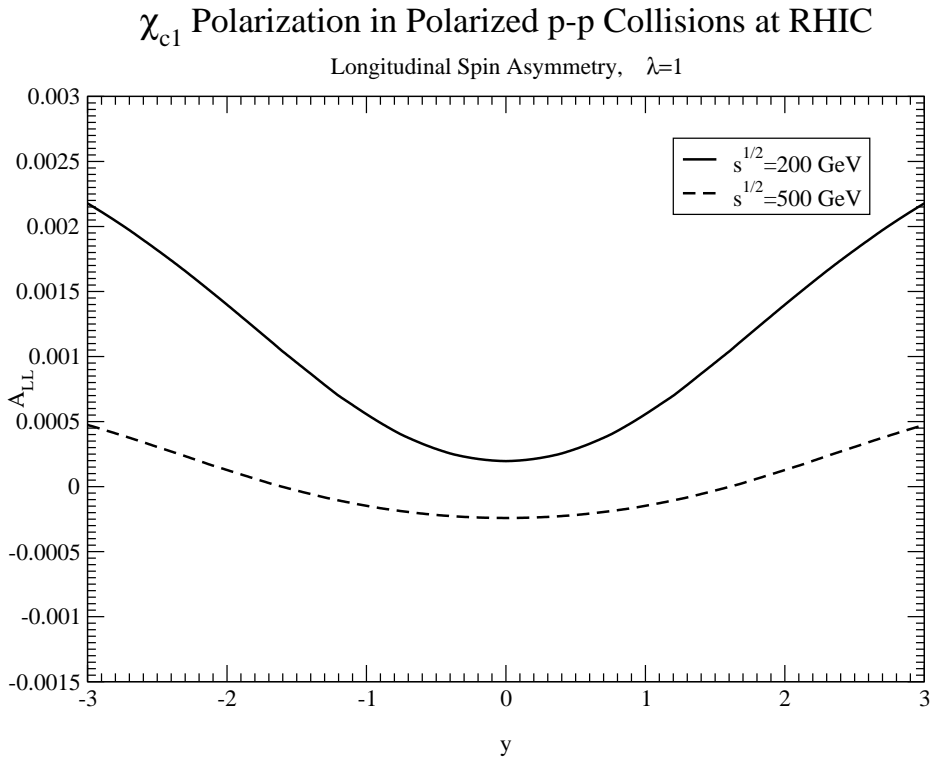}}}
\caption{ Rapidity distribution of longitudinal spin asymmetry $A_{LL}$
of $\chi_{c1}$ production at RHIC in polarized pp collisions.}
\label{fig14}
\end{figure}

In Fig. 7 we present the rapidity differential distributions
for $\chi_{c1}$ production in unpolarized p-p collisions
at $\sqrt s$ = 500 GeV. We find that the cross section
for $\chi_{c1}$ production in unpolarized p-p collisions at $\sqrt{s}$=500
GeV is larger than that at $\sqrt{s}$=200 GeV. This is due to the
enhancement of parton distribution function.

In Fig. 8 we present the rapidity differential distributions
for $\chi_{c1}$ production in polarized p-p collisions
at $\sqrt s$ = 500 GeV. Note that the values in some rapidity ranges
become negative which is due to the polarized quark distribution
function at this center of mass energy. We find that the cross section
for $\chi_{c1}$ production in polarized p-p collisions at $\sqrt{s}$=500
GeV is smaller than that at $\sqrt{s}$=200 GeV. This is due to the
polarized parton distribution function.

In Fig. 9 we present the rapidity differential distributions
for $\chi_{c2}$ production in unpolarized p-p collisions
at $\sqrt s$ = 200 GeV. The solid and dashed lines correspond to
$\lambda$=1 and 0 respectively. Note that for $\lambda$=0 the cross
section becomes small because the color octet contribution vanishes,
see eq. (\ref{unpolfin2}). The color singlet contribution is from gluon
fusion process at LO, see eq. (\ref{unpolfin2}).

In Fig. 10 we present the rapidity differential distributions
for $\chi_{c2}$ production in polarized p-p collisions
at $\sqrt s$ = 200 GeV. The solid and dashed lines correspond to
$\lambda$=1 and 0 respectively. For $\lambda$=0 the cross
section becomes small because the color octet contribution vanishes,
see eq. (\ref{eq10}). The color singlet contribution is from gluon
fusion process at LO, see eq. (\ref{polfin2}).

In Fig. 11 we present the rapidity differential distributions
for $\chi_{c2}$ production in unpolarized p-p collisions
at $\sqrt s$ = 500 GeV. The solid and dashed lines correspond to
$\lambda$=1 and 0 respectively. We find that the cross section
for $\chi_{c2}$ production in unpolarized p-p collisions at $\sqrt{s}$=500
GeV is larger than that at $\sqrt{s}$=200 GeV. This is due to the
enhancement of parton distribution function.

In Fig. 12 we present the rapidity differential distributions
for $\chi_{c2}$ production in polarized p-p collisions
at $\sqrt s$ = 500 GeV. The solid and dashed lines correspond to
$\lambda$=1 and 0 respectively. We find that the cross section
for $\chi_{c2}$ production in polarized p-p collisions at $\sqrt{s}$=500
GeV is smaller than that at $\sqrt{s}$=200 GeV. This is due to the
polarized parton distribution function.

\subsection{Spin Asymmetry of $\chi_c$ at RHIC }

In Fig. 13 we present the rapidity distributions of the
longitudinal spin asymmetry $A_{LL}$ for $\chi_{c0}$ production in
polarized p-p collisions at RHIC. The solid line is for $\sqrt{s}$=200
GeV polarized pp collisions and the dashed line is for $\sqrt{s}$=500
GeV polarized pp collisions. Note that the spin asymmetry is decreased
for higher energies. The spin asymmetry for $\chi_{c0}$
production is almost same for $\lambda$=1 and 0.

\begin{figure}[htb]
\vspace{2pt}
%\centering{\rotatebox{270}{\epsfig{figure=higgspm2.ps,height=7cm}}}
\centering{{\epsfig{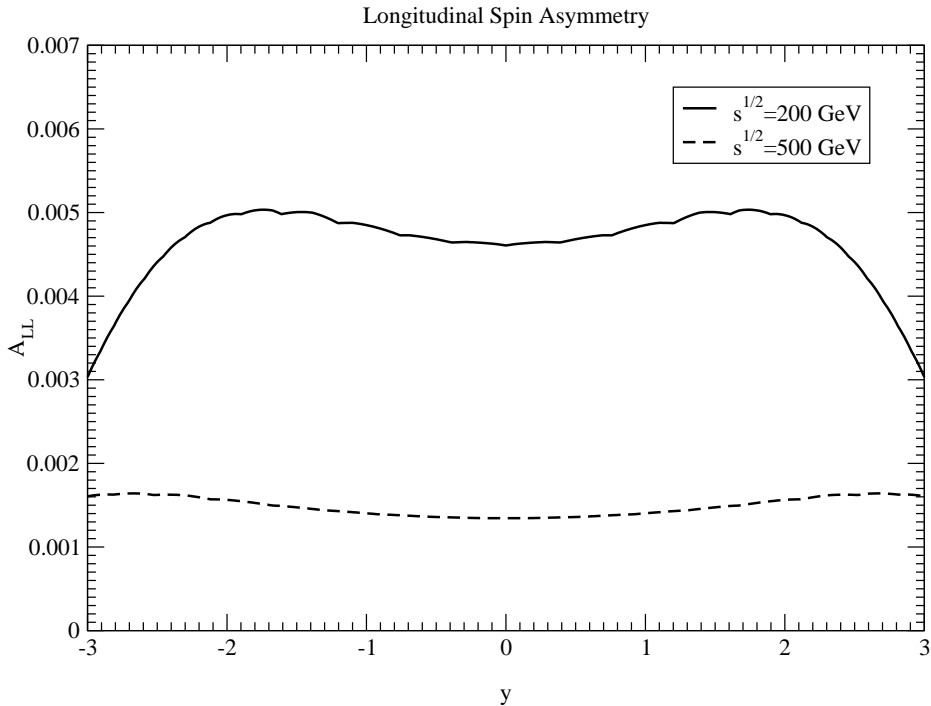}}}
\caption{ Rapidity distribution of longitudinal spin asymmetry $A_{LL}$
of $\chi_{c2}$ production at RHIC in polarized pp collisions.}
\label{fig15}
\end{figure}

In Fig. 14 we present the rapidity distributions of the
longitudinal spin asymmetry $A_{LL}$ for $\chi_{c1}$ production in
polarized p-p collisions at RHIC. The solid line is for $\sqrt{s}$=200
GeV polarized pp collisions and the dashed line is for $\sqrt{s}$=500
GeV polarized pp collisions. Note that the spin asymmetry is decreased
for higher energies. The spin asymmetry for $\chi_{c1}$
is for $\lambda$=1 which at LO arises from the color octet contribution from
quark-antiquark fusion processes, see eq. (\ref{eq10}). For $\sqrt{s}$=500 GeV the spin
asymmetry becomes negative in the rapidity range $y\sim $ 0 which is due to
the polarized quark distribution function at this center of mass energy.

In Fig. 15 we present the rapidity distributions of the
longitudinal spin asymmetry $A_{LL}$ for $\chi_{c2}$ production in
polarized p-p collisions at RHIC. The solid line is for $\sqrt{s}$=200
GeV polarized pp collisions and the dashed line is for $\sqrt{s}$=500
GeV polarized pp collisions. Note that the longitudinal spin asymmetry is decreased
for higher energies. The spin asymmetry of $\chi_{c2}$
production is almost same for $\lambda$=1 and 0.

One can see from the above figures that the cross sections for the
$\lambda = \pm 1$ states dominate over the $\lambda$=0 states.
This is explained by the fact that the coefficient in front
of the $<{\cal{O}}^{\chi_{c0}}_8(^3S_1)>$ term in eqn. (\ref{eq10})
in the color octet channel vanishes for $\lambda=0$. However, the
longitudinal spin asymmetry remains almost same for $\lambda$=1
and $\lambda$=0.

Measurement of $\chi_{cJ}(\lambda)$ polarizations with helicities
$\lambda = \pm$ 1 and $\lambda$=0 in polarized p-p collisions at RHIC
at the PHENIX detector can be useful to test the spin transfer process
in pQCD. This will also be useful to extract polarized gluon distribution
function inside proton.

\section{Conclusions}

We have studied inclusive $\chi_{cJ}$ production with definite polarizations in polarized
proton-proton collisions at $\sqrt{s}$ = 200 GeV and 500 GeV at RHIC by using
non-relativistic QCD (NRQCD) color-octet mechanism. We have presented results of
rapidity distribution of $\chi_{c0}$, $\chi_{c1}$ and $\chi_{c2}$ production
with specific polarizations in polarized p-p collisions at RHIC within the PHENIX
detector acceptance range. We have also presented the corresponding results for the spin
asymmetries.

The PHENIX experiment should be able to measure these spin
asymmetries of $\chi_{cJ}$ production. The study of heavy quarkonium production
with definite helicities in polarized p-p collisions is unique because
it tests the spin transfer processes in perturbative QCD.  As Tevatron
data for heavy quarkonium polarization \cite{expt} is not explained by
the color octet mechanism \cite{CDFpolarization}, the RHIC data may shed
some light along this direction.

The measurement of heavy quarkonium production with definite polarization
would also provide important information about quark-gluon plasma formation
\cite{qgp} at RHIC and LHC.

\end{document}